\documentclass[twocolumn]{aastex62} % use larger type; default would be 10pt

\usepackage{xspace}
\usepackage{graphicx}
\usepackage{natbib}
\usepackage{amsmath}
\usepackage{afterpage}

\newcommand{\hide}[1]{}

\newcommand{\lsim}{\ensuremath{\,\lesssim\,}\xspace}
\newcommand{\kms}{\ensuremath{\,{\rm km\,s^{-1}}}\xspace}

\newcommand{\kpc}{\ensuremath{\,{\rm kpc}}\xspace}

\newcommand{\degree}{\ensuremath{^\circ}\xspace}

\newcommand{\msun}{\ensuremath{\,\textnormal{M}_\odot}\xspace}     % solar mass
\newcommand{\hi}{{\rm H\,{\footnotesize I}}\xspace}
\newcommand{\hii}{{\rm H\,{\footnotesize II}}\xspace}

\shorttitle{Hydrogen RRL Emission from M51 and NGC\,628}
\shortauthors{Luisi et al.}

\begin{document}

\title{Hydrogen Radio Recombination Line Emission from M51 and NGC628}

\author[0000-0001-8061-216X]{Matteo Luisi}
\affiliation{Department of Physics and Astronomy, West Virginia University, Morgantown WV 26506, USA}
\affiliation{Center for Gravitational Waves and Cosmology, West Virginia University, Chestnut Ridge Research Building, Morgantown WV 26505, USA}

\author[0000-0001-8800-1793]{L. D. Anderson}
\affiliation{Department of Physics and Astronomy, West Virginia University, Morgantown WV 26506, USA}
\affiliation{Center for Gravitational Waves and Cosmology, West Virginia University, Chestnut Ridge Research Building, Morgantown WV 26505, USA}
\affiliation{Adjunct Astronomer at the Green Bank Observatory, P.O. Box 2, Green Bank WV 24944, USA}

\author[0000-0003-4866-460X]{T. M. Bania}
\affiliation{Institute for Astrophysical Research, Department of Astronomy, Boston University, 725 Commonwealth Ave., Boston MA 02215, USA}

\author[0000-0002-2465-7803]{Dana S. Balser}
\affiliation{National Radio Astronomy Observatory, 520 Edgemont Road, Charlottesville VA 22903-2475, USA}

\author[0000-0003-0640-7787]{Trey V. Wenger}
\affiliation{National Radio Astronomy Observatory, 520 Edgemont Road, Charlottesville VA 22903-2475, USA}
\affiliation{Astronomy Department, University of Virginia, P.O.~Box 400325, Charlottesville, VA 22904-4325, USA}

\author[0000-0002-3227-4917]{Amanda A. Kepley}
\affiliation{National Radio Astronomy Observatory, 520 Edgemont Road, Charlottesville VA 22903-2475, USA}

\begin{abstract}
We report the discovery of hydrogen radio recombination line (RRL) emission from two galaxies with star formation rates (SFRs) similar to that of the Milky Way: M51 and NGC\,628. We use the Green Bank Telescope (GBT) to measure $\sim 15$ Hn$\alpha$ recombination transitions simultaneously and average these data to improve our spectral signal-to-noise ratio. We show that our data can be used to estimate the total ionizing photon flux of these two sources, and we derive their SFRs within the GBT beam: $\Psi_{\rm OB} = 3.46$\msun\,yr$^{-1}$ for M51 and $\Psi_{\rm OB} = 0.56$\msun\,yr$^{-1}$ for NGC\,628. Here, we demonstrate that it is possible to detect RRLs from normal galaxies that are not undergoing a starburst with current instrumentation and reasonable integration times ($\sim 12$\,hr for each source). We also show that we can characterize the overall star-forming properties of M51 and NGC\,628, although the GBT beam cannot resolve individual \hii\ region complexes. Our results suggest that future instruments, such as the Square Kilometre Array and the Next Generation Very Large Array, will be able to detect RRL emission from a multitude of Milky Way-like galaxies, making it possible to determine SFRs of normal galaxies unaffected by extinction and to measure global star formation properties in the local universe.
\end{abstract}

%-------------------------------------------------------------------------------------------

\section{Introduction}
Radio recombination lines (RRLs) are powerful tools for studying the physical properties of the warm ($\sim 10^4$\,K), ionized gas associated with high-mass star formation. Compared to optical and near-infrared emission lines, such as H$\alpha$, RRLs have the advantage of being essentially free of extinction. Their disadvantage is reduced intensity, which restricts RRL detections to gas with higher emission measure than that traced by H$\alpha$. RRL observations have been used extensively to study Galactic \hii\ regions \citep[e.g.,][]{Bania2010,Luisi2016}, their surrounding photo-dissociation regions \citep[e.g.,][]{Wyrowski2000,Roshi2014}, and the diffuse ionized gas known as the Warm Ionized Medium that pervades the Galactic plane \citep[e.g.,][]{Roshi2000,Liu2013,Luisi2017}.  

Most studies of star formation in external galaxies focus on emission in the optical and infrared. Many young star-forming regions, however, are embedded in clouds of dust and molecular gas, with large optical extinction. RRL observations can detect emission from heavily obscured star-forming regions. These data provide information on the dynamical state of the ionized gas, and the present-day star formation rate \citep[SFR; e.g.,][]{Kepley2011}. Furthermore, by observing RRLs at different frequencies, it is possible to constrain the density and filling factor in extragalactic star-forming environments since higher-frequency transitions trace gas at higher densities \citep[e.g.,][]{Zhao1997}.

\begin{deluxetable*}{lcccccc}[htp]
\tabletypesize{\footnotesize}
\tablecaption{Observed Sources}
\tablehead{Source  & \colhead{RA} & \colhead{Dec} & \colhead{$V_0$} & \colhead{Distance} & \colhead{Type} & \colhead{SFR}\\
  & \colhead{(J2000)} & \colhead{(J2000)} & \colhead{(km\,s$^{-1}$)} & \colhead{(Mpc)} & \colhead{} & \colhead{(\msun\,yr$^{-1}$)} }
\startdata
  M51 (NGC\,5194) & 13:29:52.7 & 47:11:42 & \phn 461$^{[1]}$ & \phn 8.58$^{[3]}$ & SABb & 3.4$^{[7]}$\phn\\
  NGC\,628         & \phn 1:36:41.8 & 15:47:00 & \phn 658$^{[1]}$ & 10.19$^{[4]}$ & SAc & 0.68$^{[8]}$\\  
  M100 (NGC\,4321) & 12:22:54.9 & 15:49:20 & 1574$^{[1]}$ & 14.32$^{[5]}$ & SABb & 2.61$^{[8]}$\\
  M101 (NGC\,5457) & 14:03:12.6 & 54:20:57 & \phn 238$^{[2]}$ & \phn 6.70$^{[5]}$ & SABc & 2.33$^{[8]}$\\
  NGC\,3184        & 10:18:16.9 & 41:25:28 & \phn 583$^{[1]}$ & 11.7$^{[6]}$\phn & SABc & 0.66$^{[8]}$
\enddata
\tablenotetext{\,}{References. --- [1] \citet{Courtois2009}; [2] \citet{Makarov2011}; [3] \citet{McQuinn2016}; [4] \citet{Jang2014}; [5] \citet{Freedman2001}; [6] \citet{Jones2009}; [7] \citet{Calzetti2005}; [8] \citet{Kennicutt2011}.}
\label{tab:info}
\end{deluxetable*}

There have been numerous observations of RRL emission from nearby external galaxies using the National Radio Astronomy Observatory (NRAO) Very Large Array \citep[VLA; e.g.,][]{Seaquist1985,Anantharamaiah1993,Zhao1996,Roy2005}, and the Jansky Very Large Array \citep[JVLA; e.g.,][]{Kepley2011,Balser2017}. These studies, however, were limited to the central regions of bright starburst galaxies. Unlike RRL emission originating from the Milky Way, RRL emission from nearby galaxies is faint, with line widths greater than $\sim 100$\kms. Detection of extragalactic RRLs therefore not only requires high sensitivity and stable bandpasses, but also instrumentation with sufficiently large bandwidths that sample the entire line width. These bandwidth requirements limited extragalactic RRL observations with the VLA to frequencies $\lsim 8$\,GHz, although more recent JVLA observations do not share the same restrictions \citep{Kepley2011}.

Here we use the Green Bank Telescope (GBT) to observe galaxies in RRL emission at C-band (4--8\,GHz). With a total collecting area similar to that of the VLA, the advantage of using the GBT for these observations is
its better sensitivity for sources with extended emission. In addition, its recently upgraded C-band receiver and backend allow us to measure 22 Hn$\alpha$ transitions simultaneously. By averaging these transitions, we can increase our sensitivity considerably compared to previous RRL observations. The five galaxies in our sample are well-studied and exhibit ongoing star formation, but are not starburst galaxies. For example, the galaxy with the largest SFR in our sample, M51, has estimated SFRs ranging from 2.56\msun\,yr$^{-1}$, derived from the 20\,cm radio continuum \citep{Schuster2007}, to 5.4\msun\,yr$^{-1}$, derived from H$\alpha$ emission \citep{Kennicutt2003}, although a value of $\sim 3.4$\msun\,yr$^{-1}$ is generally accepted \citep{Calzetti2005}. Despite the relatively large total SFR, its SFR surface density of 0.015\msun\,yr$^{-1}$\,kpc$^{-2}$ is about 10--100 times lower than some of the ``weakest" starbursts \citep{Calzetti2005}. The overall star formation efficiency of M51 ($\sim 1$\%) is similar to that of the Milky Way \citep{Thronson1988}.

\section{Observations}
We made pointed C-band total power spectral observations with the GBT using the same setup as in \citet{Anderson2018}. We employed position switching with On- and Off-source integration times of 6 minutes per scan. The Off-source scans tracked the same azimuth and zenith angle path as the On-source scans, but were offset in RA such that they followed the same path on the sky. We tuned to 64 different frequencies at two polarizations within the 4--8\,GHz receiver bandpass, using 23 MHz sub-bands of 8192 channels each. Of these 64 tunings, 22 are Hn$\alpha$ transitions from $n = 95$ to $n = 117$, 25 are Hn$\beta$ lines, 8 are Hn$\gamma$ lines, and 9 are molecular lines. We only use the Hn$\alpha$ transitions for further analysis since the selected galaxies are too faint to be detected in higher-order transitions or molecular lines given our GBT configuration and integration times. 

As in our previous studies \citep[e.g.,][]{Luisi2016,Anderson2018}, we calibrated the intensity scale of our spectra using noise diodes fired during data acquisition. By observing the primary flux calibrator 3C286 we confirmed that this calibrates the data to within 10\%. In addition, we periodically observed the Galactic \hii\ regions W3 and W43 as test sources to verify the RRL intensity calibration scale and found agreement at the 10\% level with the results of \citet{Balser2011}. The GBT C-band gain to convert from antenna temperature to flux density is 2\,K\,Jy$^{-1}$ at these frequencies \citep{Ghigo2001}.

\begin{figure*}[htp]
\centering
\begin{tabular}{cc}
\includegraphics[width=.42\textwidth]{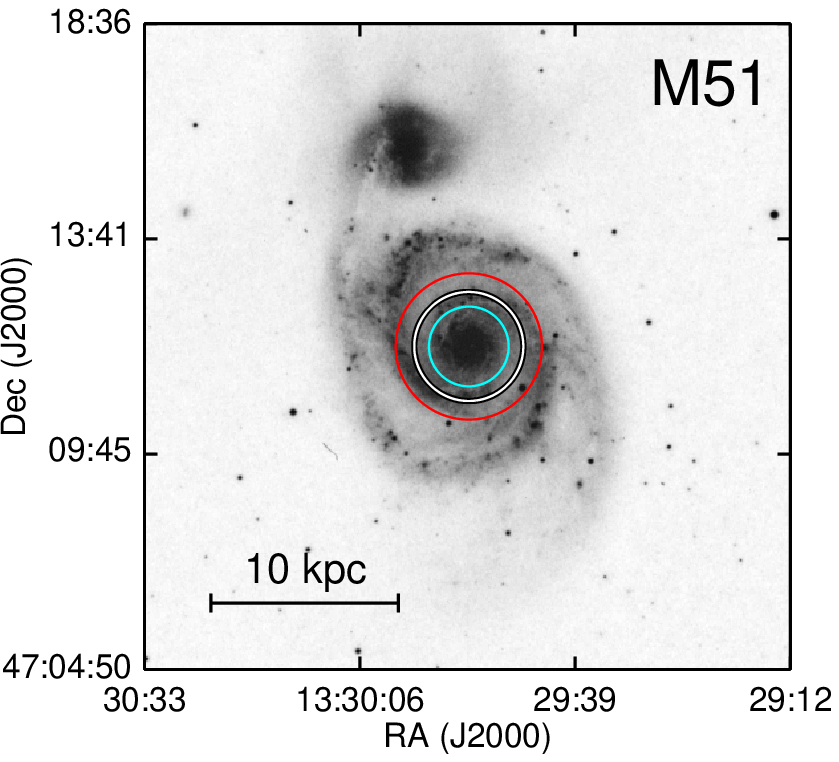} &
\includegraphics[width=.54\textwidth]{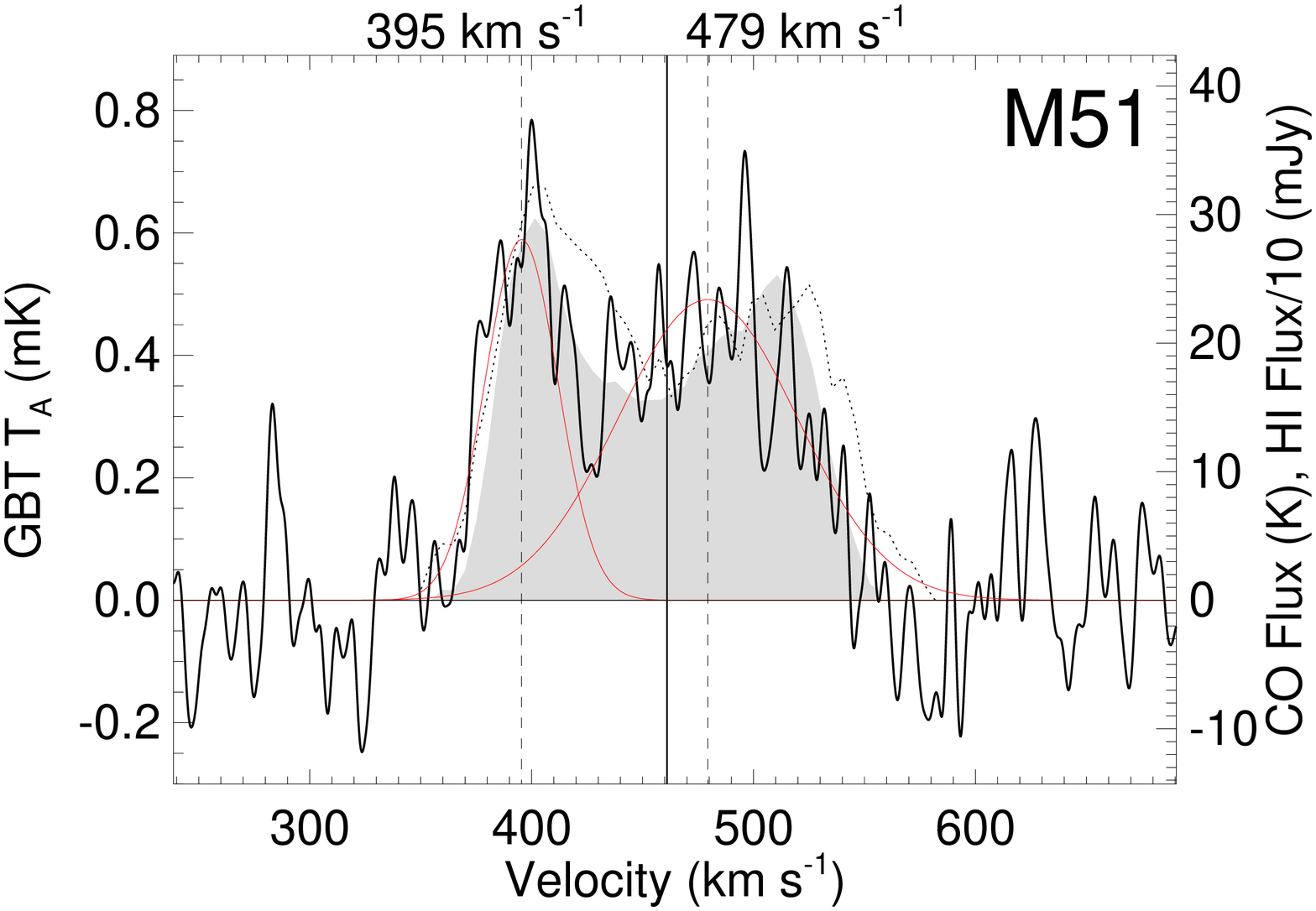}\vspace{-15pt}\\
\includegraphics[width=.42\textwidth]{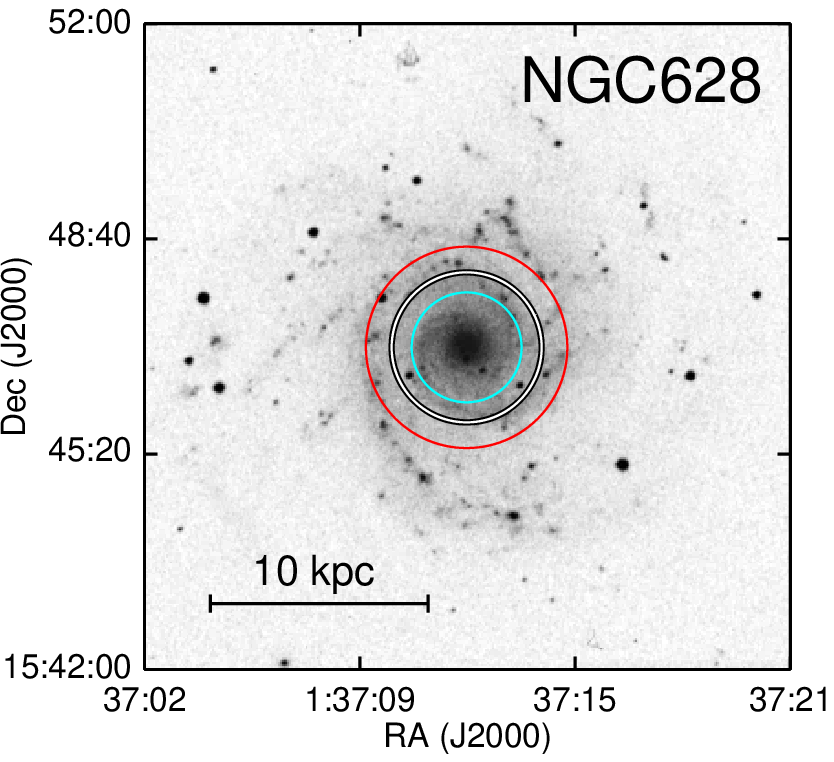} &
\includegraphics[width=.54\textwidth]{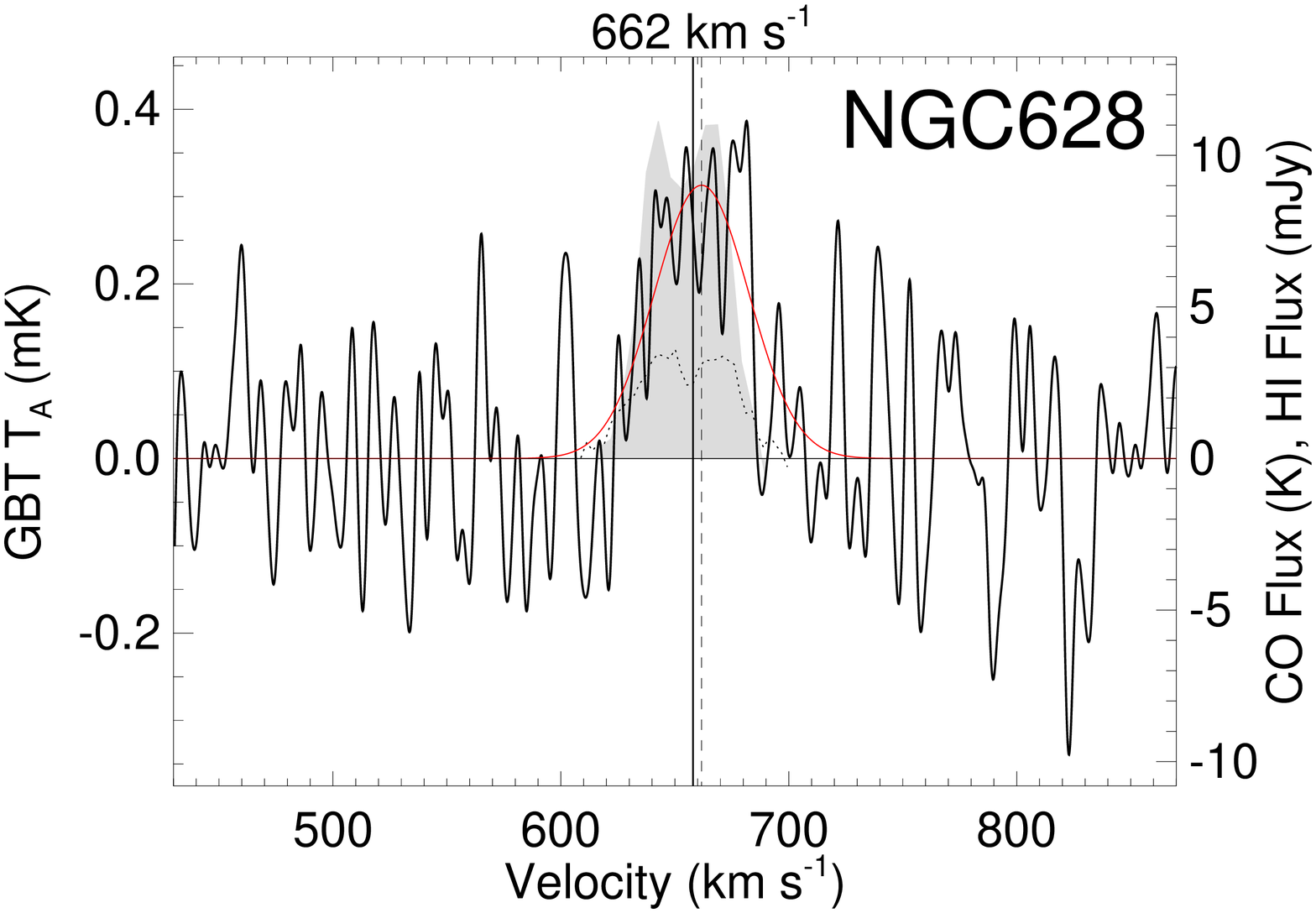}\vspace{-10pt}
\end{tabular}
\caption{\underline{Left column:} Observed galaxies. The circles are centered on the observed directions and their sizes are that of the average GBT beam (white circle with black border; 141\arcsec), the GBT beam at the highest observed frequency (7550\,MHz; light blue circle; 98\arcsec), and the GBT beam at the lowest observed frequency (4050\,MHz; red circle; 183\arcsec). The background images were taken from the 645\,nm STScI Digitized Sky Survey and the scale bars are derived from the distances given in Table~\ref{tab:info}. \underline{Right column:} RRL spectra of the observed galaxies, smoothed to a spectral resolution of 4.07\kms. Plotted is the GBT antenna temperature as a function of heliocentric velocity. The recession velocities of the galaxies are given by the solid vertical lines. We detect hydrogen emission above ${\rm S/N} = 5$ for M51 and NGC\,628, which we approximate with the Gaussian model fits shown in red. The centers of the Gaussian peaks are indicated by the dashed vertical lines; we show the corresponding velocities at the top of the figure. For comparison, we show HERACLES CO data integrated over the GBT beam as the shaded gray regions. Shown with the dotted lines are integrated \hi\ data from the THINGS survey (M51, M101, NGC\,628, and NGC\,3184) and the VIVA survey (M100).\label{fig:spectra}}
\end{figure*}
\renewcommand{\thefigure}{\arabic{figure}}

\renewcommand{\thefigure}{\arabic{figure} (cont.)}
\addtocounter{figure}{-1}
\begin{figure*}[htp]
\centering
\begin{tabular}{cc}
\includegraphics[width=.42\textwidth]{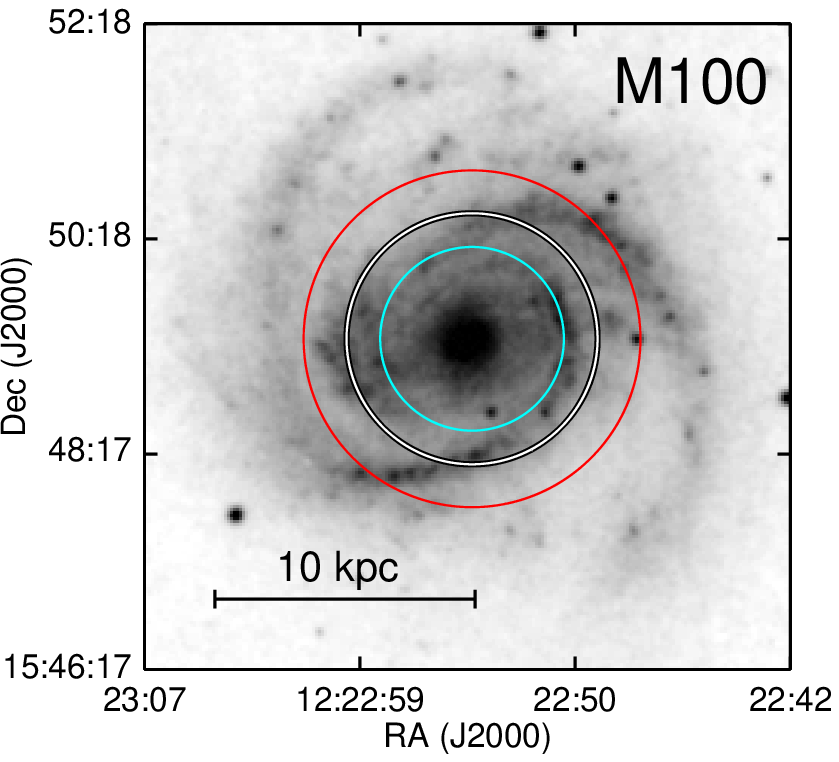} &
\includegraphics[width=.54\textwidth]{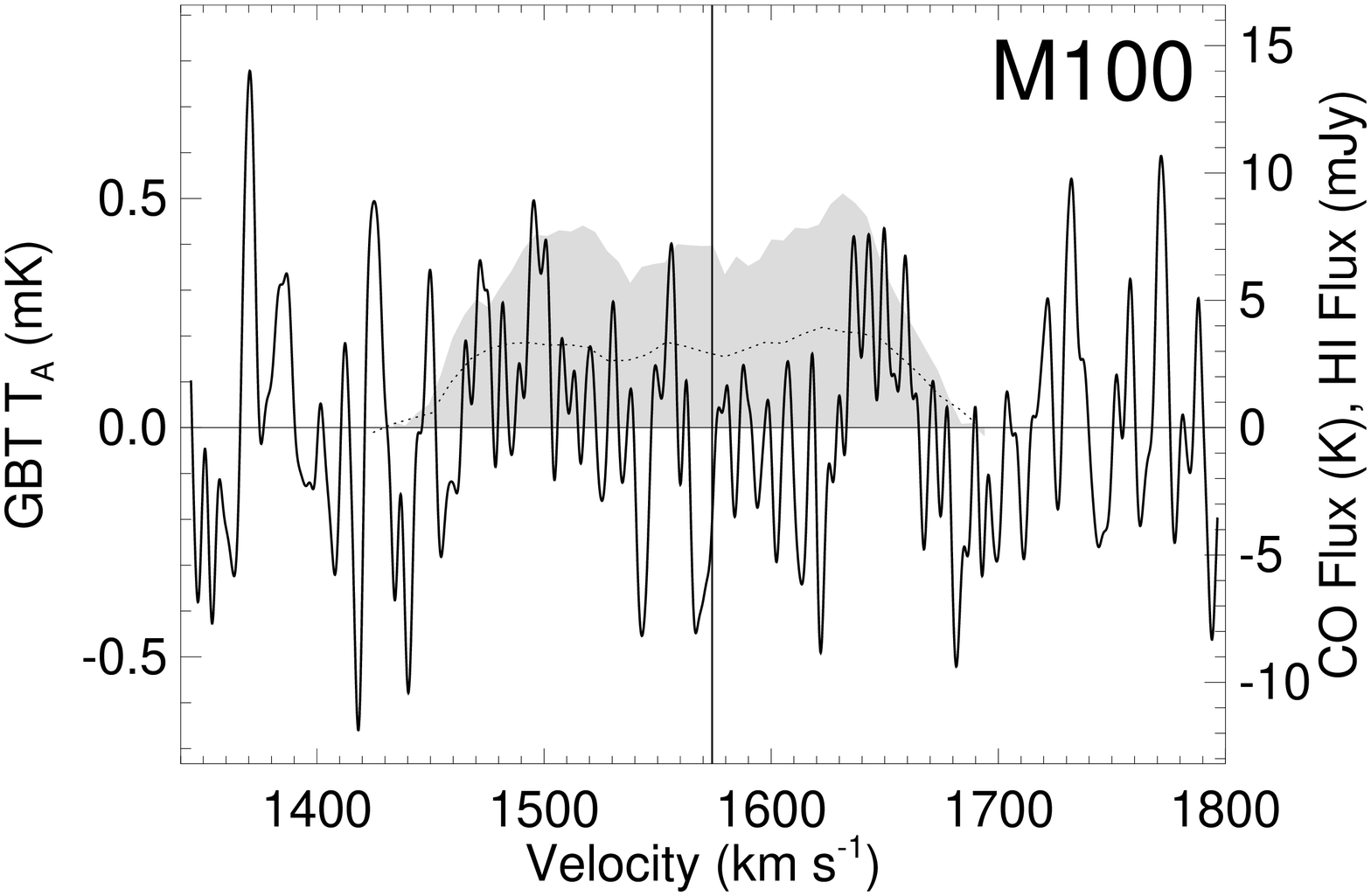}\vspace{-15pt}\\
\includegraphics[width=.42\textwidth]{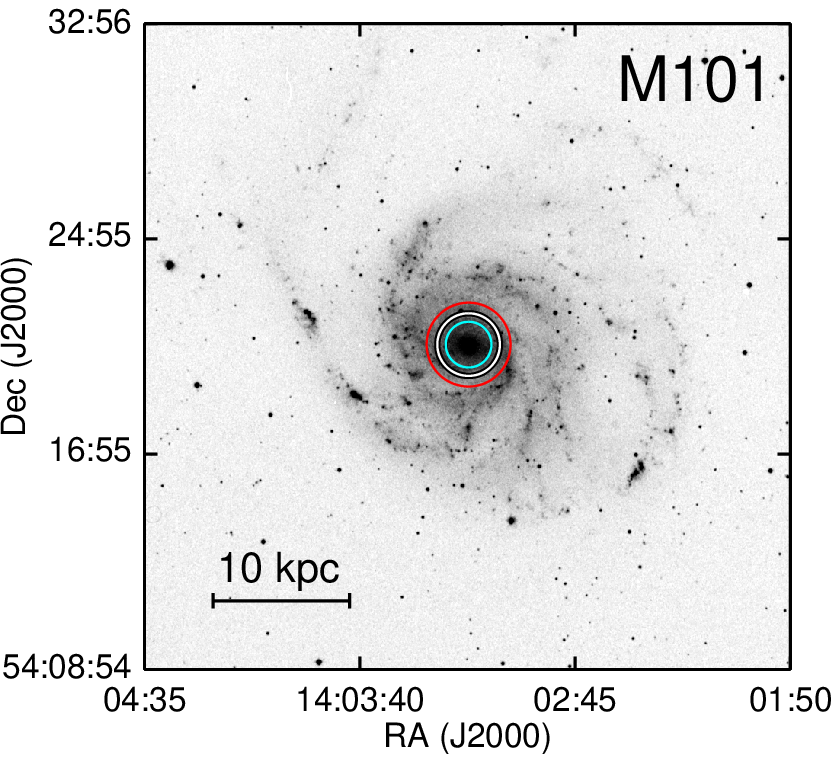} &
\includegraphics[width=.54\textwidth]{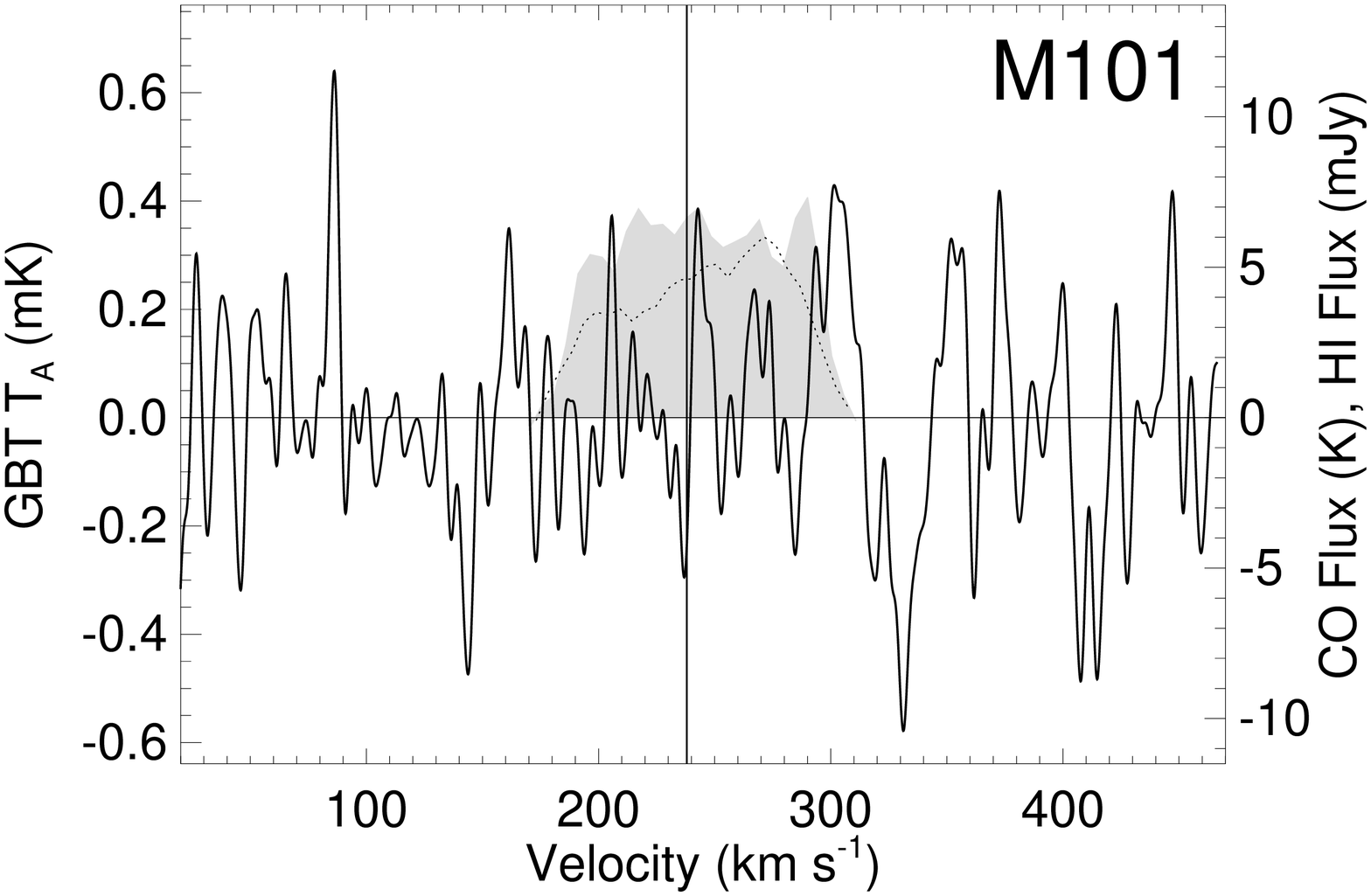}\vspace{-15pt}\\
\includegraphics[width=.42\textwidth]{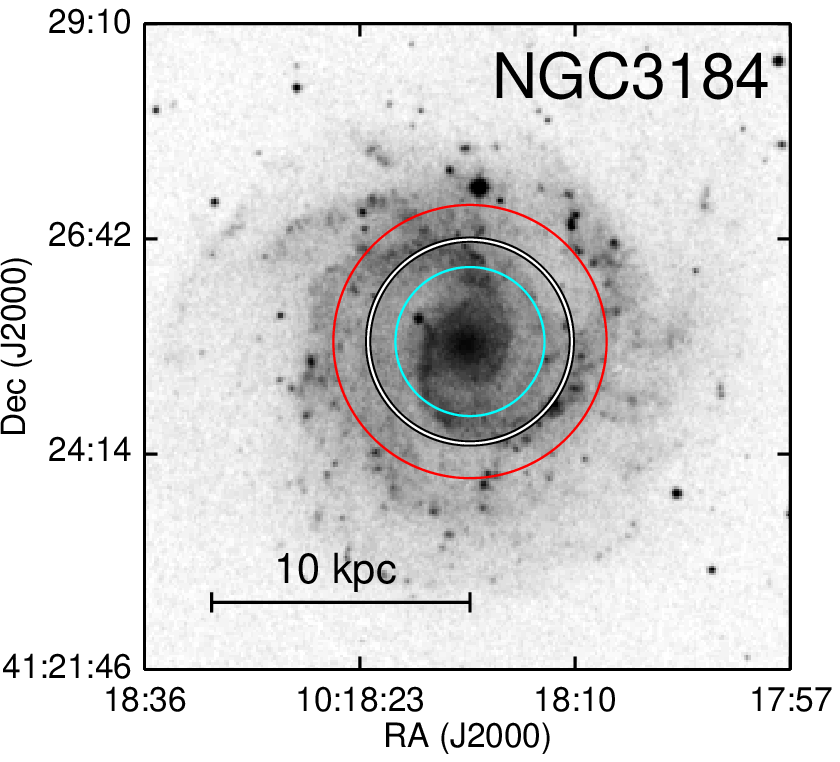} &
\includegraphics[width=.54\textwidth]{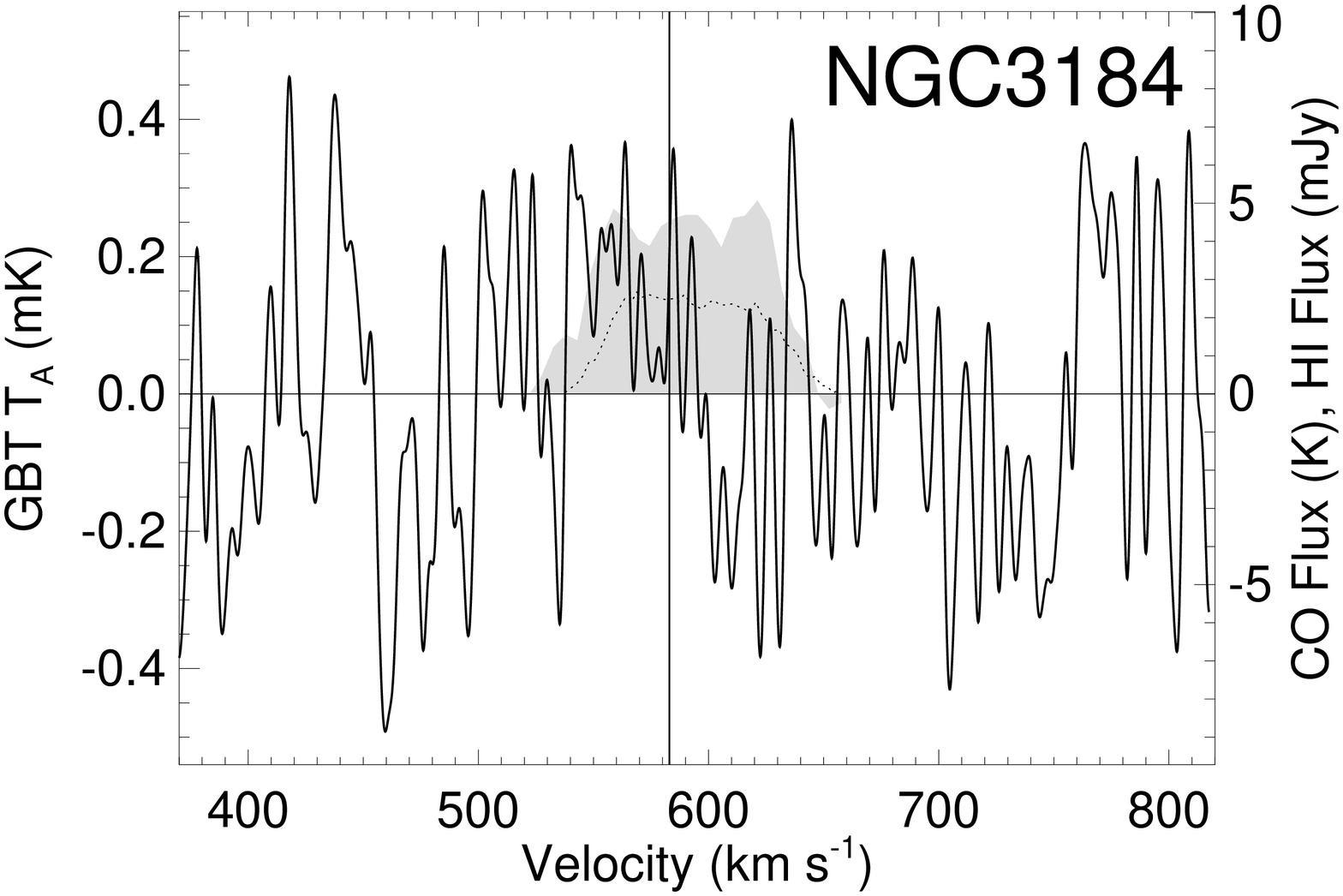}\vspace{-10pt}\\
\end{tabular}
\caption{}
\end{figure*}
\renewcommand{\thefigure}{\arabic{figure}}
\renewcommand{\thefigure}{\arabic{figure} (cont.)}

Our sample included five face-on galaxies: M51, NGC\,628, M100, M101, and NGC\,3184. We pointed the telescope toward the center of each galaxy to ensure that a large fraction of the galaxy's ionized gas is located within the GBT beam (see Figure~\ref{fig:spectra}, left column). Due to the large average GBT half power beam width (HPBW) of 141\arcsec, we cannot resolve individual star-forming complexes. In addition, the HPBW varies from 98\arcsec\ to 183\arcsec\ across our frequency range, and we therefore sample slightly different portions of the galaxy for each transition. Our total integration times, $t_{\rm intg}$, for each source range from 3.6 to 12.8 hours. In Table~\ref{tab:info} we list the sources, the coordinates of the observed directions, the mean Heliocentric recession velocities of the galaxies based on \hi\ measurements ($V_0$), their distances, their Hubble types, and their reported SFRs.

We use TMBIDL\footnote{V8.0, see https://github.com/tvwenger/tmbidl.git.} to reduce and analyze our RRL data \citep[see][]{Bania2014}. For each observed direction, we discard spectra affected by radio frequency interference (RFI). Out of the 22 Hn$\alpha$ lines, three are consistently affected by RFI and we never use them for further analysis (namely the H114$\alpha$, H115$\alpha$, and H117$\alpha$ transitions). Six lines are sometimes affected by RFI (H95$\alpha$, H96$\alpha$, H97$\alpha$, H98$\alpha$, H104$\alpha$, and H116$\alpha$), and the remaining 13 lines are almost never affected. We re-grid the $\sim 15$ Hn$\alpha$ lines unaffected by RFI to the velocity resolution of the H95$\alpha$ data and shift the spectra so that they are aligned in velocity \citep{Balser2006}. For M100, several interpolated spectra were affected by baseline instabilities, reducing the number of good Hn$\alpha$ lines to $\sim 10$. We then average the spectra using a weighting factor of $t_{\rm intg} \, T_{\rm sys}^{-2}$ where $t_{\rm intg}$ is the integration time and $T_{\rm sys}$ is the system temperature. After removing a fourth-order baseline from the averaged spectrum, we smooth the spectrum to a velocity resolution of 4.07\kms.

We define the signal-to-noise ratio (S/N) of the detected hydrogen lines using the method described by \citet{Lenz1992},
\begin{equation}
{\rm S/N} = 0.7 \left( \frac{T_{\rm L}}{\rm rms} \right) \left( \frac{\Delta V}{\lambda} \right) ^{0.5},
\end{equation}
where $T_{\rm L}$ is the peak line intensity, rms is the root-mean-squared spectral noise, $\Delta V$ is the full width at half maximum (FWHM) line width, and $\lambda = 4.07$\kms is the FWHM of the Gaussian smoothing kernel.

We fit Gaussian models to all averaged hydrogen RRLs with a S/N of at least 5, from which we derive the peak line intensities, their FWHM values, and heliocentric velocities. We summarize our results in Table~2, which lists the source, the line intensity, the FWHM line width, the velocity, the rms noise in the spectrum, the S/N, and the total integration time for each direction, including the corresponding 1$\sigma$ uncertainties of the Gaussian fits.

\section{Results}
Our observational setup allows us to detect, for the first time, hydrogen RRL emission from external galaxies with SFRs comparable to that of the Milky Way. We detect RRL emission from two of the five observed sources: the grand design spirals M51 and NGC\,628. 

M51 (NGC\,5194) is a mostly face-on ($i = 33$\degree) SABb galaxy, with a mass of $1.6 \times 10^{11}$\msun\ \citep{Holmberg1965}, located at a distance of $8.58 \pm 0.14$\,Mpc \citep{McQuinn2016}. \citet{Bell1978} observed M51 at 6\,GHz using the Algonquin Radio Observatory, but failed to detect the H102$\alpha$ line, citing an upper limit of 8.2\,mJy. More recently, \citet{Aladro2015} performed a 3-mm survey of nearby galaxies but did not detect RRL emission from M51.

We show our RRL spectrum for M51 in Figure~\ref{fig:spectra}. We find that the detected emission is well-approximated by two Gaussian models of similar height centered at 396\kms and 479\kms, respectively. We also show for comparison CO spectral line data from the `HERA CO-Line Extragalactic Survey' \citep[HERACLES; see][]{Leroy2009}, integrated over the size of the average GBT beam. There is good morphological agreement between the RRL spectrum of M51 and the HERACLES data, suggesting that we recover RRL emission from the same regions traced by CO. The morphology of the RRL spectrum is also comparable to neutral hydrogen emission data from `The \hi\ Nearby Galaxy Survey' \citep[THINGS;][see Figure~\ref{fig:spectra}]{Walter2008}, integrated over the GBT beam.

We use the method described by \citet{Heiles1992} to derive the ionization rate of M51 using our RRL data. We estimate the total recombination rate of M51 assuming local thermodynamic equilibrium,
\begin{equation}
\frac{\dot{N}_{\rm H}}{\int T_{\rm b} \, d \nu} \approx 8.1 \times 10^7 \, T_3^{0.8},
\end{equation}
where $\dot{N}_{\rm H}$ is the total recombination rate per cm$^{2}$, $\int T_{\rm b} \, d \nu$ is the frequency-integrated brightness temperature in kHz\,K, and $T_3$ is the temperature of the ionized gas in $10^3$\,K. We assume here that the RRL emitting region is extended evenly across the GBT beam, in which case the observed antenna temperature, $T_{\rm A}$, can be set equal to $\eta_{\rm MB}T_{\rm b}$, where $\eta_{\rm MB}$ is the main beam efficiency of the GBT. Here, $\eta_{\rm MB} \approx 0.94$, assuming an aperture efficiency of 0.7 at 6\,GHz \citep{Maddalena2010,Maddalena2012}.

\begin{deluxetable*}{lccccccccc}
\tabletypesize{\footnotesize}
\tablecaption{Hydrogen RRL Parameters}
\tablehead{Source  & \colhead{$T_{\rm L}$} & \colhead{$\sigma T_{\rm L}$} & \colhead{$\Delta V$} & \colhead{$\sigma \Delta V$} & \colhead{$V$} & \colhead{$\sigma V$}  & \colhead{rms} & \colhead{S/N} &\colhead{$\rm t_{\rm intg}$}\\
  & \colhead{(mK)} & \colhead{(mK)} & \colhead{(km\,s$^{-1}$)} & \colhead{(km\,s$^{-1}$)} & \colhead{(km\,s$^{-1}$)} & \colhead{(km\,s$^{-1}$)} & \colhead{(mK)} & \colhead{} & \colhead{(min)}}
\startdata
  M51      & 0.589 & 0.008 & 38.7 & 0.7 & 395.4 & 0.3 & 0.114 & 11.2 & 684 \\
            & 0.491 & 0.005 & 95.0 & 1.6 & 479.4 & 0.6 & 0.114 & 14.6 & \\
  NGC\,628  & 0.313 & 0.006 & 48.7 & 1.3 & 661.7 & 0.5 & 0.112 & 6.8 & 768 \\            
  M100      & \nodata & \nodata & \nodata & \nodata & \nodata & \nodata & 0.234 & \nodata & 216 \\
  M101      & \nodata & \nodata & \nodata & \nodata & \nodata & \nodata & 0.208 & \nodata & 228 \\
  NGC\,3184 & \nodata & \nodata & \nodata & \nodata & \nodata & \nodata & 0.229 & \nodata & 216 
\enddata
\label{tab:hii}
\end{deluxetable*}

Using our average GBT HPBW of 141\arcsec, assuming a distance of 8.58\,Mpc to M51, and setting the recombination rate equal to the ionizing photon flux per second, $N_{\rm Lyc}$, we find
\begin{equation}
N_{\rm Lyc} \approx 3.9 \times 10^{52} \, {\rm s}^{-1} \times T_3^{0.8}.
\end{equation}
Here we make the assumption that both dust attenuation and escape of photons into the intergalactic medium are negligible. We assume an electron temperature of $7000$\,K typical of star-forming regions at small galactocentric radii \citep[e.g.,][]{Balser2011} and find an ionizing flux, $N_{\rm Lyc} \approx 1.87 \times 10^{53}$\,s$^{-1}$, for M51. We estimate the SFR using
\begin{equation}
N_{\rm Lyc} = 5.4 \times 10^{52} \, {\rm s}^{-1} \times \Psi_{\rm OB},
\end{equation}
where $\Psi_{\rm OB}$ is the SFR averaged over the lifetime of OB stars in \msun\,yr$^{-1}$ \citep{Anantharamaiah2000}. We find $\Psi_{\rm OB} = 3.46$\msun\,yr$^{-1}$ for M51 within the GBT beam which agrees with the value of 3.4\msun\,yr$^{-1}$ reported by \citet{Calzetti2005}. 

The largest uncertainty contributions to $\Psi_{\rm OB}$ are the assumed value of $T_3$ and residual baseline frequency structure, which can have a significant effect on $\int T_{\rm b} \, d \nu$. Based on the magnitude of observed variations of the baseline, we estimate that these contributions may change $\Psi_{\rm OB}$ by up to 30\%. 

The polynomial order of the subtracted baseline only has a small effect on the uncertainty in $\int T_{\rm b} \, d \nu$. While low-order polynomial baselines may not adequately fit the line-free portion of the receiver bandpass, higher-order baselines may introduce artificial structures that can affect the measured peak line intensities and FWHM values. To quantify this effect, we recalculate the integrated line intensity for M51 after removing polynomial baselines of order one to six. We find an average root-mean-squared deviation in $\int T_{\rm b} \, d \nu$ of 4\% between the six baseline models, indicating that the baseline removal does not have a major effect on our derived parameters. We include this contribution in our uncertainty estimate of $\Psi_{\rm OB}$.

We note that if our assumption of extended emission across the GBT beam is inaccurate, the calculated $\Psi_{\rm OB}$ becomes a lower limit, since in that case $T_{\rm b} \geq T_{\rm A}/\eta_{\rm MB}$. In addition, the beam only covers the central $\sim 3$\kpc of the galaxy, and therefore our value of $\Psi_{\rm OB}$ underestimates the total SFR of M51. We summarize our results in Table~\ref{tab:nly}, which lists the source, the integrated line intensity, $\int T_{\rm A} \, dV$, the ionizing flux, $N_{\rm Lyc}$, and the SFR averaged over the lifetime of OB stars, $\Psi_{\rm OB}$.

\begin{deluxetable}{lccc}
\tabletypesize{\footnotesize}
\tablewidth{0pt}
\tablecaption{Derived Parameters}
\tablehead{Source  & \colhead{$\int T_{\rm A} \, dV$} & \colhead{$N_{\rm Lyc}$} & \colhead{$\Psi_{\rm OB}$}\\
  & \colhead{(mK\kms)} & \colhead{($10^{52}$\,s$^{-1}$)} & \colhead{(\msun\,yr$^{-1}$)} }
\startdata
  M51 & \phn \phn 69.8 & \phn \phn 18.7 & \phn \phn 3.46\\
  NGC\,628 & \phn \phn 15.2 & \phn \phn \phn 3.0 & \phn \phn 0.56\\  
  M100 & $< 39.6$ & $< 29.6$ & $< 5.47$ \\
  M101 & $< 30.3$ & \phn $< 4.9$ & $< 0.92$ \\
  NGC\,3184 & $< 27.5$ & $< 13.7$ & $< 2.54$ 
\enddata
\tablecomments{We assume a 1$\sigma$ uncertainy of $\pm 30$\% for all derived parameters.}
\label{tab:nly}
\end{deluxetable}

NGC\,628 is almost completely face-on ($i \lsim 10$\degree) and similar in morphology to M51, albeit with an SFR of only 0.68\msun\,yr$^{-1}$, derived from the combination of H$\alpha$ and 24\,$\mu$m data \citep{Kennicutt2011}. We detect RRL emission from NGC\,628, centered at 662\kms (see Figure~\ref{fig:spectra}). As for M51, there is good morphological agreement between our RRL spectrum, the CO data from the HERACLES survey and the \hi\ data from the THINGS survey. Assuming a distance of 10.19\,Mpc \citep{Jang2014}, we estimate the ionizing photon flux of NGC\,628 using Eq.~2 and find $N_{\rm Lyc} \approx 3.0 \times 10^{52}$\,s$^{-1}$ within the GBT beam. This corresponds to $\Psi_{\rm OB} = 0.56$\msun\,yr$^{-1}$ (Eq.~4). We again calculate the effect that the polynomial order of the subtracted baseline has on the uncertainty in $\int T_{\rm b} \, d \nu$. We find a deviation of 10\% between the baseline models of order one to six and estimate a total uncertainty contribution of $\pm 30$\% for $\Psi_{\rm OB}$.

A more comprehensive method to estimate the electron density and ionizing photon flux from RRL observations has been proposed by \citet{Anantharamaiah1993}.  This method models a collection of individual \hii\ regions, each characterized by an electron temperature, electron density, linear size, and turbulent velocity. Unfortunately, we cannot use this method, since precise knowledge of the line-to-continuum ratio is required. While we are able to roughly constrain the continuum temperature, $T_{\rm C}$, for two galaxies (M51 and M101) from our data, the uncertainties in $T_{\rm C}$ are large ($\gg 20\%$) for these low-intensity sources.

We do not detect hydrogen RRL emission from the other three galaxies in our sample (M100, M101, and NGC\,3184; see Figure~\ref{fig:spectra}). We also show for comparison HERACLES CO spectral line data and THINGS \hi\ data for these sources. Since M100 was not included in the THINGS survey, we compare this source with \hi\ data from the `VLA Imaging of Virgo in Atomic gas' (VIVA) survey instead \citep{Chung2009}. 

We estimate upper limits for the ionizing photon flux and the SFR for each source undetected in RRL emission. We fit a Gaussian profile to the HERACLES CO data integrated over the GBT beam and calculate $T_{\rm L}$ expected for RRL emission with ${\rm S}/{\rm N} = 5$ based on the FWHM of the CO line and our spectral rms (see Eq.~1). We then use this value to find upper limits on $\int T_{\rm A} \, dV$, $N_{\rm Lyc}$, and $\Psi_{\rm OB}$ (see Table~\ref{tab:nly}). It is surprising that we do not detect RRL emission from M101 given its relatively large SFR of 2.33\msun\,yr$^{-1}$ \citep{Kennicutt2011}. Since the GBT beam only covers the central 2.3\,kpc of the galaxy, we speculate that we are missing a significant amount of star formation at larger galactocentric radii \citep[see, e.g.,][]{Grammer2014}.

With the increased sensitivity of future radio observatories it will become possible to study the ionized gas in a large number of external galaxies, determine extragalactic SFRs unaffected by extinction, and measure global star formation properties in the local universe. The Square Kilometre Array (SKA), the Next Generation Very Large Array (ngVLA), and the Five-hundred-meter Aperture Spherical Radio Telescope (FAST) will be able to detect RRL emission from a multitude of Milky Way-like galaxies. With a total collecting area of $\sim 10^6$\,m$^2$ and a spectral bandwidth of 300\,MHz at L-band, the SKA will be able to observe $\sim 10$ Hn$\alpha$ lines simultaneously. The ngVLA will have an effective collecting area of $\sim 4 \times 10^4$\,m$^2$ and an instantaneous bandpass of 8.7\,GHz at C- and X-band, covering $\sim 40$ Hn$\alpha$ lines. Its spectral rms at this frequency is 81.7\,$\mu$Jy\,beam$^{-1}$ before averaging the lines, assuming a 1\arcsec\ spatial resolution, an integration time of 1\,hr and a 10\kms channel width \citep{Selina2017}. With a collecting area of $\sim 7 \times 10^4$\,m$^2$, FAST will have an L-band sensitivity of $\sim 18$\,K\,Jy$^{-1}$ and system temperatures of $\sim 20$\,K \citep{Li2016}. These instruments should be able to detect RRL emission from high-mass star-forming regions within external galaxies up to distances of several tens of Mpc. We show that even if individual \hii\ region complexes within these galaxies remain unresolved, their overall star-forming properties can be characterized.

\section{Conclusions}
We used the GBT to search for $\sim 6$\,GHz hydrogen recombination line emission from five external face-on galaxies. We detected hydrogen RRL emission from two of the sources, M51 and NGC\,628. M51 is similar to the Milky Way in both SFR and star formation efficiency \citep{Calzetti2005,Thronson1988}. We estimate the ionizing photon flux from our spectral line data and find $N_{\rm Lyc} \approx 1.87 \times 10^{53}$\,s$^{-1}$, assuming an electron temperature of $7000$\,K. This corresponds to an SFR of $\Psi_{\rm OB} = 3.46$\msun\,yr$^{-1}$, which is comparable to the value of 3.4\msun\,yr$^{-1}$ reported by \citet{Calzetti2005}. For NGC\,628, we find $N_{\rm Lyc} \approx 3.0 \times 10^{52}$\,s$^{-1}$ and $\Psi_{\rm OB} = 0.56$\msun\,yr$^{-1}$.

Our study shows that it is possible to detect RRLs from normal galaxies that are not undergoing a starburst with current instrumentation and in reasonable integration times. We highlight the importance of simultaneously observing multiple RRL transitions that can be averaged together to increase the overall sensitivity. We also show that the ionizing fluxes and SFRs of external galaxies can be estimated from the GBT RRL emission data alone, although the high uncertainty in the measured continuum temperature and the large beam size make it necessary to use simplistic models such as the one reported by \citet{Heiles1992}. The good agreement between our results and the literature, however, suggests that the \citet{Heiles1992} model gives reasonable estimates of the ionizing photon flux and SFR for these sources. 

\acknowledgements 
We thank the anonymous referee for insightful comments that improved the clarity of this manuscript. We thank West Virginia University for its financial support of GBT operations, which enabled the observations for this project. This work made use of THINGS, `The HI Nearby Galaxy Survey' \citep{Walter2008} and HERACLES, `The HERA CO-Line Extragalactic Survey' \citep{Leroy2009}. The Digitized Sky Survey was produced at the Space Telescope Science Institute under U.S.~Government grant NAG W-2166. The images of these surveys are based on photographic data obtained using the Oschin Schmidt Telescope on Palomar Mountain and the UK Schmidt Telescope. The plates were processed into the present compressed digital form with the permission of these institutions.

\textit{Facility:} Green Bank Telescope.
\textit{Software:} TMBIDL \citep{Bania2014}.
%-------------------------------------------------------------------------------------------

\bibliographystyle{aasjournal}

\end{document}